\documentstyle[11pt,nyas]{article}

\def\th{\thinspace}

\input{epsf.sty}

\begin{document}

\def\la{\mathrel{\mathpalette\fun <}}
\def\ga{\mathrel{\mathpalette\fun >}}
\def\fun#1#2{\lower3.6pt\vbox{\baselineskip0pt\lineskip.9pt
\ialign{$\mathsurround=0pt#1\hfill##\hfil$\crcr#2\crcr\sim\crcr}}}

\newcommand{\al}{\alpha} 
\renewcommand{\b}{\beta} 
\newcommand{\de}{\delta}
\newcommand{\De}{\Delta}
\newcommand{\ep}{\epsilon}
\newcommand{\Ga}{\Gamma}
\newcommand{\ka}{\kappa}
\newcommand{\io}{\iota}
\newcommand{\La}{\Lambda}
\newcommand{\Om}{\Omega}
\newcommand{\om}{\Omega_m}
\newcommand{\si}{\sigma}
\newcommand{\Si}{\Sigma}
\newcommand{\th}{\theta}
\newcommand{\vth}{\vartheta}
\newcommand{\vph}{\varphi}
\newcommand{\ra}{\rightarrow}
\newcommand{\XX}{\mbox{$\cal X$}}
\newcommand{\tr}{\mbox{tr}}
\newcommand{\hor}{\mbox{hor}}
\newcommand{\grad}{\mbox{grad}}
\newcommand{\lap}{\triangle}
\newcommand{\arctg}{\mbox{arctg}}
\newcommand{\bm}[1]{\mbox{\boldmath $#1$}}
\newcommand{\vs}{{v_{{\scriptscriptstyle 12}}}}
\newcommand{\bapm}{{b_{{\scriptscriptstyle A}}}}
\newcommand{\xl}{{\xi^{{\scriptscriptstyle (1)}}}} 
\newcommand{\xxl}{{\xi_{{\scriptscriptstyle 1}}}} 
\newcommand{\xq}{{\xi^{{\scriptscriptstyle (2)}}}} 
\newcommand{\xxq}{{\xi_{{\scriptscriptstyle 2}}}} 
\newcommand{\se}{{\sigma_{{\scriptscriptstyle 8}}}} 
\renewcommand{\baselinestretch}{1.01}
\newcommand{\be}{\begin{equation}}
\newcommand{\ee}{\end{equation}}
\newcommand{\gsim}{\stackrel{>}{\sim}}
\newcommand{\lsim}{\stackrel{<}{\sim}}
\newcommand{\bea}{\begin{eqnarray}}
\newcommand{\eea}{\end{eqnarray}}
\newcommand{\bean}{\begin{eqnarray*}}
\newcommand{\eean}{\end{eqnarray*}}
\newcommand{\dd}{\partial}
\newcommand{\x}{\vec x}
\newcommand{\r}{\vec r_{12}}
\newcommand{\xb}{\bar{\xi}}
\newcommand{\xbb}{\bar{\hspace{-0.08cm}\bar{\xi}}}
\newcommand{\xbl}{{\xb^{{\scriptscriptstyle (1)}}}} 
\newcommand{\xbq}{{\xb^{{\scriptscriptstyle (2)}}}} 
\newcommand{\lu}{\,h^{-1}{\rm kpc}}
\newcommand{\Mlu}{\,h^{-1}{\rm Mpc}}
\newcommand{\RoA}{{{\rho}_{{\scriptscriptstyle A}}}}
\newcommand{\Rone}{{{\rho}_{{\scriptscriptstyle 1}}}}
\newcommand{\Rotwo}{{{\rho}_{{\scriptscriptstyle 2}}}}
\newcommand{\rone}{{{\vec r}_{{\scriptscriptstyle 1}}}}
\newcommand{\ro}{\scriptscriptstyle {\rho}}
\newcommand{\rtwo}{{{\vec r}_{{\scriptscriptstyle 2}}}}
\newcommand{\rA}{{{\vec r}_{{\scriptscriptstyle A}}}}
\newcommand{\va}{{{\vec v}_{{\scriptscriptstyle A}}}}
\newcommand{\vv}{{{\vec v}_{{\scriptscriptstyle 12}}}}
\newcommand{\vJ}{{{v^J}_{12}}}
\newcommand{\vone}{{{\vec v}_{{\scriptscriptstyle 1}}}}
\newcommand{\vtwo}{{{\vec v}_{{\scriptscriptstyle 2}}}}
\newcommand{\vab}{{{\left[ v_{r}(\rone)-v_{r}(\rtwo) \right]}}}
\newcommand{\xig}{{\xi_{{\scriptscriptstyle {\rm g}}}}}
\newcommand{\vsg}{{v_{{\scriptscriptstyle 12{\rm g}}}}}
\newcommand{\da}{{{\delta}_{{\scriptscriptstyle A}}}}
\newcommand{\done}{{{\delta}_{{\scriptscriptstyle 1}}}}
\newcommand{\dtwo}{{{\delta}_{{\scriptscriptstyle 2}}}}
\newcommand{\dg}{{\delta_{{\scriptscriptstyle {\rm g}}}}}
\newcommand{\dgA}{{\delta_{{\scriptscriptstyle {\rm g}A}}}}

\title{The transition to nonlinearity and new constraints on biasing}
\author{Roman Juszkiewicz$^{1,2,3}$ \& Enrique Gazta\~naga$^{4,5}$
\\
$^{1}$ J. Kepler Astronomical Center, 65-265 Zielona G{\'o}ra, Poland \\
$^{2}$ D{\'e}partement de Physique Th{\'e}orique, Universit{\'e}
de Gen{\`e}ve, CH-1211 Gen{\`e}ve, Switzerland \\
$^{3}$ N. Copernicus Astronomical Center, 00-716 Warsaw, Poland \\
$^{4}$ INAOE, Astrofisica, Tonantzintla, 
Apdo Postal 216 y 51,  Puebla 7200, Mexico \\
$^{5}$ Institut d'Estudis Espacials de Catalunya,  
IEEC/CSIC, Edf. Nexus-201 - c/ Gran Capitan 2-4, 08034 Barcelona, Spain}


\begin{abstract} 

We present two new dynamical tests of the biasing hypothesis.  The
first is based on the amplitude and the shape of the galaxy-galaxy
correlation function, $\xi_g(r)$, where $r$ is the separation of the
galaxy pair. The second test uses the mean relative peculiar velocity
for galaxy pairs, $\vs(r)$. This quantity is a measure of the rate of
growth of clustering and it is related to the two-point
correlation function for the matter density fluctuations, $\xi(r)$.
Under the assumption that galaxies trace the mass ($\xi_g = \xi$), the
expected relative velocity can be calculated directly from the
observed galaxy clustering.  The above assumption can be tested
by confronting the expected $\vs$ with direct measurements from
velocity-distance surveys. 
Both our methods are checked against N-body experiments and then compared
with the $\xi_g(r)$ and $\vs$ estimated from the
{\sc APM} galaxy survey and the Mark III catalogue, respectively.  Our
results suggest that cosmological density parameter is low, $\Omega_m
\approx 0.3$, and that the {\sc APM} galaxies trace the mass at
separations $r \ga 5 \Mlu$, where $h$ is the Hubble constant in units
of 100 km s$^{-1}$Mpc.  The present results agree with earlier
studies, based on comparing higher order correlations in the {\sc APM}
with weakly non-linear perturbation theory. Both approaches constrain
the linear bias factor to be within $20\%$ of unity. If the existence
of the feature we identified in the {\sc APM} $\xi_g(r)$ -- the
inflection point near $\xi_g = 1$ -- is confirmed by more accurate
surveys, we may have discovered gravity's smoking gun: the long
awaited ``shoulder'' in $\xi$, generated by gravitational dynamics and
predicted by Gott and Rees 25 years ago.


\end{abstract}


\bigskip

\section{Introduction}

We have been recently working on two projects, directly related
to the topic of this Meeting -- the transition to nonlinearity
in gravitational clustering. We have therefore decided to present
some of our results here. The first idea -- to use the inflection 
point in the galaxy-galaxy
correlation function to constrain
biasing is in a more mature state than the second,
which uses measurements of the relative motions in pairs of galaxies.  
Accordingly, the first project is almost a paper -- it has been
submitted for publication in {\it Monthly Notices}. The
results of our work on the second project are still far from
completion but ripe enough to be presented and discussed.

\subsection{The CDM crisis and biasing}

The concept that galaxies may not be fair tracers of the mass distribution
was introduced in the early eighties, 
in part in response to the observation that
galaxies of different morphological types have different spatial
distributions, hence they cannot all trace the mass
(there are two excellent reviews on the subject:
Strauss \& Willick 1995 and Hamilton 1998). However,
there was also another reason: to ``satisfy the theoretical
desire for a flat universe'' (Davis et al. 1985, p.391).
More precisely, biasing was introduced to
reconcile the observations with the predictions of the Einstein-de Sitter
cold dark matter (CDM) dominated model.
At the time, it seemed that just a simple rescaling of the
overall clustering amplitude by setting $\xi_g(r) = b^2\xi(r)$,
where $b \approx 2$ might do the job (Davis et al. 1985).
However, very soon thereafter, it became clear that this is not enough:
while the unbiased ($b = 1$) $\xi(r)$ had too large an amplitude
at small $r$, the biased model did not have enough
large-scale power to explain the observed bulk motions
(Vittorio et al. 1987). A similar conclusion could be drawn
form comparison of the relative amplitude of clustering 
on large and small scales (eg Maddox et al 1990).
The problem with the shape of $\xi(r)$
became explicit when measurements of $\xi_g(r)$ 
showed that the optically selected galaxies follow an
almost perfect power law over nearly three orders of magnitude
in separation. This result disagrees with 
N-body simulations. The standard ($\Omega_m = 1$)
CDM model as well as its various modifications, including 
$\Omega_m < 1$ and a possible non-zero cosmological constant, 
fail to match the observed power law
(see Fig 11-12 in Gazta\~naga 1995, Jenkins et al. 1998; most of these
problems were already diagnosed by Davis et al. 1985). 
Two alternative ways out of this impasse were
recently discussed by Rees (1999) and Peebles (1999).
We believe that it will be helpful to discuss both approaches
because their existence provides the motivation for our work.

\subsection{Environmental cosmology}

A possible response to the CDM crisis is to build a model where
simple phenomena, like the power-law 
behavior of $\xi_g$ are much more complicated than they seem. 
In particular, one can explore the possibility that the
emergence of large scale structure is not driven by gravity alone
but by ``environmental cosmology'' -- a complex
mixture of gravity, star formation and dissipative hydrodynamics
(Rees 1999). A phenomenological formalism, appropriate
for this approach was recently proposed Dekel \& Lahav (1999).
According to the old, ``linear biasing'' prescription, at each
smoothing scale, the galaxy- and the dark matter-density fields,
$\delta_g$ and $\delta$, are 
related  by the linear function
\begin{equation}
\delta_g \;= \; b\,\delta \; ,
\end{equation} 
where $b$ is a 
time- and scale-independent constant. In the new picture,
the relationship between the two fields is neither linear nor
deterministic. The biasing factor, here defined as the ratio
of the rms values of the two fields,
$b = \sigma_g/\sigma$ is allowed to depend on the smoothing scale    
as well as on the cosmological time. A convenient measure of
the stochasticity in the relationship between the two fields
is the cross-correlation coefficient, 
\begin{equation}
R \; \equiv \;{\langle\delta\delta_g\rangle
\over \sigma_g\sigma} \; ; \; \;\; |R| \leq 1 \; .
\label{bullshit_coefficient}
\end{equation}
The special case $R = 1$ describes the deterministic bias,
while $R = 0$ corresponds to the ``maximum stochasticity'',
or the case when the two fields are
completely uncorrelated. The parameters $R$ and $b$ describe
he biasing and stochasticity in the galaxy distribution
relative to the spatial mass distribution. When referring to
specific galaxy surveys, we will sometimes use subscripts, e.g. 
$b_{IRAS}$ for the $IRAS$ survey. These quantities should
be distinguished from the bias and stochasticity measures 
for two classes of galaxies 
of different morphological types, e.g. for early (subscript $e$)
and late (subscript $l$) galaxies: 
$\, b_{el} \equiv \sigma_e/\sigma_l$, and
$\,R_{el} \, \equiv \,\langle\delta_e\delta_l\rangle(\sigma_e\sigma_l)^{-1}$.

The above parameters can be estimated or constrained by more or
less direct measurements from galaxy redshift surveys, peculiar
velocity data and other observations. They can be also studied
in semi-analytic theoretical models (Seljak 2000, Scoccimarro et al.
2000), hydrodynamic simulations or semi-analytic models combined 
with N-body simulations (eg. Blanton et al. 2000, 
Somerville et al. 2001 and references therein).

\subsection{Constraints on biasing}

If biasing is indeed important and complicated,
we should expect that $b, b_{el}, R$ and $R_{el}$ are all significantly
different from unity and scale-dependent. As we show below,
at sufficiently large scales, $r \ga 10 h^{-1}$Mpc,
there is actually evidence to the
contrary: the admissible deviations of $b$ and $R$ are small
and always comparable to the accuracy of the measurements. 

{\bf Skewness and higher moments.}
The strongest constraints on large (weakly non-linear) scales
come from the measurements of the two-,
three- and four-point connected moments of the density field in the
{\sc APM} catalogue. These constraints are obtained as follows. 
One assumes that galaxies
trace the mass and the large-scale structure we observe today grew out
of small-amplitude, Gaussian density fluctuations in an expanding,
self-gravitating non-relativistic gas. If our assumption is correct, 
by now nonlinear gravitational
instability would have driven the distribution away from
gaussianity, generating skewness and higher connected
moments, which can be calculated analytically
(Juszkiewicz et al. 1993, Bernardeau 1994a, 1994b).
The assumptions about initial gaussianity and lack of biasing
can then be tested by comparing the analytical predictions
with observations. The predictions are in
good agreement with the data from the APM (Gazta\~naga 1994,
Gazta\~naga \& Frieman 1994, Frieman \& Gazta\~naga 1999)
as well as the IRAS
PSCz catalogue (Feldman et al. 2000) and suggest that
$b(r)$ is within $20\%$ of unity for $r \ga 10 h^{-1}$Mpc,
or at linear scales (where the clustering amplitude is
less than unity).

{\bf Redshift distortions.}
Some of the most radical   
claims that $b_{el}$ can be as large as 1.5 to 2 are 
based on comparisons of the
estimates of the strength of clustering in the 1.2 Jy 
IRAS catalogue with optical redshift surveys 
(Strauss \& Willick 1995 and references therein). 
Indeed, those earlier studies,
summarized by Hamilton (1998), gave a redshift distortion
parameter $\, \beta_{IRAS} = \Omega_m^{0.6}/b_{IRAS} =
0.77 \pm 0.22$, while the average and the standard deviation
of the same parameter, obtained from optical surveys is,
according to the same compilation, $\, \beta_{optical} =
0.52 \pm 0.26$, implying a relative bias $\, b_{optical}/b_{IRAS} =
b_{el} \approx 1.5$. The 
most recent analysis (Hamilton et al. 2000)
of the larger PSCz sample \cite{will}
gives $\beta_{IRAS} = 0.41 \pm 0.13$, consistent with
$\, b_{el} = 1$. Moreover, Hamilton et al. also conclude that
their results are consistent with $R_{IRAS}(r) \approx 1$
at $r \ga 10 h^{-1}$Mpc. More recently Hamilton and Tegmark (2000)
have studied the large as well as small-scale clustering
in real space, reconstructed from the redshift space data
of the PSCz survey and found evidence
for scale-dependent bias; however, the effect is
significant at small, strongly-nonlinear scales only.

{\bf Large scale flows.}
Recent measurements of the mean relative pairwise velocity of
galaxies allow an independent estimate of $\Omega_m$ and the biasing
parameter.  These
results are consistent with $\, \Omega_m \approx 0.3$
and $R \approx b \approx
1, \; b_{el} \approx R_{el}
\approx 1$ at separations $r \ga 5 h^{-1}$Mpc \cite{rj00}.  

{\bf Weak lensing.}
The values of $b$ and $\Omega_m$, derived from large scale
flows are consistent with recent measurements
of cosmic shear correlations based on the VIRMOS deep
imaging survey (Van Waerbeke et al. 2001). On much smaller
scales, $r \approx 1 h^{-1}$Mpc, weak lensing data from
the Sloan Survey give $\; \Omega_m R/b \approx 0.3$ (Fischer et
al. 1999), again in agreement with the estimates from the pairwise motions,
although unlike the relative velocity approach, the Sloan results
suffer from degeneracies between $\Omega_m$ and bias.

{\bf Qualitative arguments.}
Qualitatively, strong biasing effects
would be difficult to reconcile with the well
known fact that the $L_*$ galaxies, dwarf galaxies and
IRAS galaxies have strikingly similar distributions, all
avoiding the voids (Peebles 1993, pp. 638-642).
Another empirical argument against biasing is provided by the
universal character of the observed Tully-Fisher and fundamental plane
relations (see, for example Binney 1999; Peebles 1999).

{\bf Simulations.} All simulations of the galaxy formation process,
either of hydrodynamic or semi-analytic variety predict 
morphological segregation as well as
$b$ and $R$ parameters dependent on scale and time. Since
these models are at a relatively early stage of their development,
the details differ from model to model
on small scales and at high redshifts, i.e. exactly where their
$b$ and $R$ parameters are significantly different from unity
and where clustering is strongly nonlinear. However,
most of these numerical experiments broadly agree that with
decreasing redshift and increasing scales, $b$ and $R$ approach unity 
(see eg. Figure 18 in Somerville et al. 2001 and references therein).

In the end, it may turn out that to explain the observed structure,
all we need is just the plain gravitational instability theory,
leaving complex non-gravitational physics on scales below a Megaparsec 
to cosmogony, directly involving star formation.  We discuss this
possibility below.

\subsection{What you get is what you see}

An obvious alternative to environmental cosmology 
was recently discussed by Peebles (1999), who pointed out that
``as Kuhn has taught us, complex interpretations
of simple phenomena have been known to be precursors of paradigm
shifts'' and perhaps after fifteen years of attempts to salvage
the CDM model with biasing, 
it is time to abandon this approach, as well as
the biasing hypothesis itself as ``another phlogiston'' (all quotations
in inverted commas are from Peebles 1999).
Instead, one can explore a simpler
option, that galaxies trace the mass distribution, or 
\begin{equation}
\xi_g  \; = \; \xi \; \; {\rm and} \; \; R \; = \; b \; = \;1 \; ,
\end{equation}
at least for local (low redshift), optically selected 
galaxies with a broad magnitude sampling.   
This approach rests on the idea that no matter how
or where galaxies form, they must eventually fall into the 
dominant gravitational wells and therefore trace the underlying mass
distribution (see Peebles 1980, hereafter
LSS; Fry 1996). Our purpose here is
to test this idea, using measurements of relative velocities
of pairs of galaxies and the shape of the two-point correlation function.

\subsection{Outline of the paper}

In this paper we propose two new tests of the biasing hypothesis,
which involve two measures of clustering. The first is
the two-point correlation function of mass density fluctuations,
$\xi$. The second is a measure of the rate of gravitational clustering
-- the relative velocity of particle pairs, $\vs$. We describe our
theoretical model in the next section. Our analytic
formulae used to test the biasing hypothesis are checked
against N-body simulations in \S3. In \S4 we apply our tests
to the {\sc APM} galaxy survey. Finally, in \S5 we summarize and discuss
our results.

\section{Theory}

\subsection{The relative velocity}

The relative pairwise velocity $\vs$ was introduced
in the context of the {\sc BBGKY} theory (Davis and Peebles 1977),
describing the dynamical evolution of a collection of particles
interacting through gravity. In this discrete picture,
$\vv$ is defined as the mean value of the peculiar
velocity difference of a particle pair at separation ${\vec r}$
({\sc LSS}, Eq. 71.4). In the fluid limit, its analogue is the
pair-density weighted relative velocity \cite{fisher94,rj98a},
\begin{eqnarray}
\vv(r) \; = \;
{{ \langle(\vone -
\vtwo ) (1 + \done)(1 +\dtwo ) \rangle}
\over {1 \; + \; \xi(r)}} \; ,
\label{def}
\end{eqnarray}
where $\, \va \,$ and $\, \da = \RoA / \langle \rho \rangle - 1 \,$
are the peculiar velocity and fractional density contrast
of matter at a point $A = 1,2,\ldots$.
The brackets $\langle \ldots \rangle$ denote ensemble averages
for pairs of points at a fixed separation
$r=|\rone - \rtwo|$,  while $\xi(r) =
\langle\done\dtwo\rangle$. In gravitational instability theory,
the magnitude of $\vv(r)$ is related to 
$\xi(r)$ through the pair conservation equation
({\sc LSS}, Eq. 71.6). 

Recently Juszkiewicz et al. (1999, hereafter {\sc JSD})
have proposed an approximate solution of the pair
conservation equation.
Using Eulerian perturbation theory, they solved 
the equation for $\vs(r)$ to second order in $\xi$.
They also proposed an
interpolation between their large-$r$ perturbative limit,
and the well known small separation limit -- the
stable clustering solution, $\vs(r) = - Hr$,
where $H$ is the Hubble constant ({\sc LSS}, \S71).
The resulting ansatz is given by 
\begin{eqnarray}
\vs (r) \; &=& \; - \, {\textstyle{2\over 3}} \, H f r \,
\xbb (r)[1 + \alpha \; \xbb (r)] \;,
\label{2nd-order}\\
\xb(r) \; &=& \; (3/r^3)\, \int_0^r \, \xi(x) \, x^2 \, dx \; 
\equiv \; \xbb(r) \, [\, 1 + \xi(r) \, ] \; .
\label{xb}
\end{eqnarray}
Here
$\alpha$ is a parameter, determined by the logarithmic slope
of $\xi(r)$, while
$f = d \, \ln D/d \ln a$, with
$D(a)$ being the standard linear growing mode
solution and $a$ -- the cosmological
expansion factor ({\sc LSS}, \S11). 
The solution (\ref{2nd-order}) assumes Gaussian initial
conditions, and the dynamics of clustering is assumed to be dominated by 
the gravity of inhomogeneities in a
pressure-less fluid of non-relativistic particles. For all such
models, including those with a non-zero cosmological constant,
$f \approx \Omega_m^{0.6}$ (Peebles 1993, \S13).
If the correlation function is given
by a pure power law, $\xi \propto r^{-\gamma}$, where
$0 < \gamma < 3$, the parameter $\alpha$
is given by 
\begin{equation}
\alpha \; \approx \; 1.2 \; - \; 0.65 \, \gamma \; .
\label{alpha}
\end{equation}
This formalism can be generalized for a non-power law $\xi(r)$
by replacing $\gamma$ in Eq.~(\ref{alpha}) with an effective slope,
\begin{equation}
\gamma_o \; \equiv \; - \, d\ln\xi/d\ln r \left|_{r_o} \right. \;
,
\label{eff}
\end{equation}
evaluated at separation $r = r_o$, defined by the condition
\begin{equation}
\xi(r_o) \; = \; 1 \; .
\label{ro}
\end{equation}
{\sc JSD} tested the equations
(\ref{2nd-order}) - (\ref{ro}) against
high resolution N-body simulations, provided by
the Virgo consortium (Jenkins et al. 1998). They
found an excellent agreement between the streaming
velocity, predicted by their ansatz and the $\vs(r)$,
measured from the simulations
in the entire dynamical range, $\,0.1 < \xi < 10^3$. 
However, the N-body experiments they used
were confined to four different
{\sc CDM}-like models, considered by Jenkins et al. (1998).
As we have already pointed out
in the Introduction,  models of this kind
fail to reproduce the observed $\xi_g(r)$
unless one resorts to a highly contrived,
scale- and time-dependent biasing function
(Gazta\~naga 1995, Jenkins et al. 1998, Peebles 1999). One of our
objectives here is to test the validity {\sc JSD} ansatz 
for $\vs(r)$ against a new set of N-body
simulations, which differ significantly
from those originally considered
by {\sc JSD}. Here we use simulations with
a more realistic $\xi(r)$, inferred by Baugh \&
Gazta{\~n}aga (1996) from the measurements of galaxy-galaxy correlations
in the {\sc APM} survey (see the description below;
from now on, we will refer to these numerical experiments
and their initial conditions as {\sc APM}-like).

\subsection{The inflection point}

In the gravitational instability theory, newly forming mass clumps
are generally expected to collapse before relaxing to virial equilibrium.
If this were so, $|\vs(r)|$ would have to be larger than the
Hubble velocity $Hr$ to make $\vs(r) + Hr$ negative. As a consequence,
the slope of $\xi$, 
\begin{equation}
d \ln \xi(r) /d \ln r \; = \; - \, \gamma(r) \; ,
\end{equation}
must increase at separations where
$\xi(r,t) \approx 1$.
This effect was recognized long ago by
Gott \& Rees (1975). When 
the expected ``shoulder'' was not found in the observed
galaxy-galaxy correlation function, Davis \& Peebles (1976)
introduced the so-called pre-virialization conjecture as
a way of reducing the size of the jump in $\gamma(r)$
(the conjecture involves non-radial motions
within the collapsing clump; see the discussion in LSS, \S 71
and Peebles 1993, pp. 535 - 541; see also Villumsen \& Davis 1986; 
{\L}okas et al. 1996 and \cite{rs96}). 
Later observational work showed a 
clear break in the shape of $\xi$ for several
redshift and angular catalogues, which was early
evidence for the linear to non-linear transition,
pointed out by Guzzo and collaborators 
(see the review by Guzzo 1997).

The arguments, raised by Peebles and Davis
(1976) were qualitative rather than quantitative. 
Quarter a century later the precision of N-body simulations as well as
the quality of the observational data have improved dramatically
enough to justify a reexamination of the problem.
The actual shape of the correlation function near $\xi = 1$
can be investigated with high resolution N-body simulations like those run
by the Virgo Consortium (Jenkins et al. 1998).  
In all four of the Virgo models, 
the slope of $\xi(r)$ exhibits a striking feature. Instead of a shoulder,
or a simple discontinuity in $\gamma(r)$,
however, $\xi(r)$ has an inflection point,  
\begin{equation}
d^2\xi(r)/dr^2 = 0 \; ,
\end{equation}
which occurs at a uniquely defined separation $r = r_*$.
At this separation, the logarithmic
slope of $\xi$ reaches a local maximum, 
$\, d \ln\xi/d\ln r  = - \gamma_*$. 
In all four of the models JSD investigated, the inflection
point indeed appears near the transition $\xi = 1$, as expected
by the earlier speculations, involving the ``shoulder'' in $\xi$.
Namely, $r_*$ is almost identical with the scale of nonlinearity:
\begin{equation}
r_* \; \approx \; r_o \; .
\label{test}
\end{equation}
More precisely, a comparison of Figure 1 in JSD with
Figure 8 in Jenkins et al. (1998) gives 
\begin{equation}
|r_o - r_*| \; < \;0.1 \,r_o
\label{precise_test}
\end{equation}
for all four considered models.
Moreover, for all models, studied by JSD, the $- \gamma$ vs. $r$ dependence
can be described as an S-shaped curve,
with a maximum at $r = r_* \approx r_o$, and a minimum
at a smaller separation. For $r \geq r_*$,
the nonlinear slope (measured
from Virgo simulations) follows the linear $\gamma(r)$,
determined by the initial conditions.
The separation $r_*$ 
is therefore also the branching point between the linear and
nonlinear $\gamma(r)$ curves, which 
are identical for $r > r_*$ 
(actually, they differ a little;
the small differences between the two curves
can be entirely attributed to sampling errors
in the N-body experiments; see Figure 1 in JSD). 
This property can be used to identify $r_*$ in noisy simulations,
and/or observations,
when the noise in the measured $\gamma(r)$ curve does not allow
a direct determination of $r_*$ as the position of the 
maximum in $- \gamma(r)$. 

If the relation  (\ref{test}) is indeed a general  
property of gravitational clustering, 
it can be used as a test of biasing as follows.
Suppose the biasing factor is a scale-independent 
constant, significantly
greater than unity: $b \gg 1$. Then 
$\xi_g \gg \xi$ and the relation (\ref{test}) will break down. 
For a power-law correlation function,
$\xi_g (r) = (r_o/r)^{\gamma} = b^2\xi(r)$, and instead
of equation (\ref{test}) we will have
\begin{equation}
r_* \; \approx \; r_o \, b^{-2/\gamma} \; .
\label{biased_test}
\end{equation}
Since the observed slope is $\gamma \approx 1.8$, 
for $b = 2$, the shoulder in the correlation function
should appear at a separation smaller than a half
of the $r_o$! The above argument can be generalized to 
a broader class of models,
allowing scale-dependent bias provided $b(r)$ is a 
smooth monotonic function and $b(r_o)$ is significantly
greater than unity. Whenever these conditions apply,
strong biasing always implies $r_o \gg r_*$. Hence,
the comparison of these two scales determined from
observations can be used as a diagnostic for biasing.

\section{N-body simulations}

\subsection{Initial conditions}

In this section, we compare our ansatz for $\vs(r)$
with the results from P$^3$M simulations. We use
{\sc APM}-like models for the initial shape of $P(k)$
in Baugh \& Gazta\~naga (1996). The
box size is $600 ~h^{-1}$ Mpc (or $300 ~h^{-1}$ Mpc) with 
$200^3$ (or $100^3$)  dark matter particles.
The {\sc APM}-like models
have Gaussian initial conditions with a  power spectrum 
inferred from correlations in the {\sc APM} survey,
following the procedure, introduced by Baugh \&
Gazta\~naga 1996. The power spectrum of the
density fluctuations is related to
$\xi(r)$ through the usual Fourier transform,
\begin{equation}
P(k) \;= \; \int \; \xi(r)\exp(i{\bf k\cdot r})\,d^3r \; .
\end{equation}
Following the prescription of Baugh \& Gazta{\~n}aga,
we assume an initial power spectrum of the form
\begin{equation}
P(k) = {A k\over \left(1+(k/0.05 h~ 
{\rm Mpc}^{-1})^2\right)^{1.6}}~.
\label{Pkapm}
\end{equation}
The linear spectrum, given above is designed to evolve into a
nonlinear one, matching the APM measurements of $P(k)$ under the
assumption of no bias. The normalization constant $A$ is directly
related to the degree of inhomogeneity of the mass distribution at the
end of the simulation, parametrized by $\se$ -- the rms mass density
contrast, measured in spheres of a radius of 8 $h^{-1}$Mpc.
Following Baugh and Gazta{\~n}aga, we choose the constant $A$ in order
to ensure that at the end of the simulation, $\se =0.85$.  The quality
of the agreement of the evolved $\xi(r)$ with the APM data at small
separations, where $\xi \gg 1$, depends on the assumed value of
$\Omega_m$. However, as we show below, for $r \geq 2 h^{-1}$Mpc and
$\xi < 3$, this effect is negligible (compare Figures 1 and 3).  This
range of separations and clustering amplitudes brackets from both
sides the region, on which we focus on here: the transition between
$\xi < 1$ and $\xi > 1$.  We use simulations with $\Lambda = 0$ and
two different values of the density parameter: $\Omega_m = 1$ and 0.3.

\subsection{Estimators}

Two different estimators for $\xi(r)$ and $\vs$ are used. To construct
the first estimator, we cut the simulation box into cubical pixels of
size $\Delta_l$, placed on a regular lattice. The density fluctuation
amplitude at the $i$th pixel is
\begin{equation}
\delta_i \; = \; {N_i \over \langle N \rangle} \, - \, 1 \; ,
\end{equation}
where $N_i$ is the particle 
count in that pixel. The estimator for the 
two-point function is then:
\be
{\hat \xi}^{(1)}(r) \; = \; {1\over N_r}\sum_{i,j} \delta_i 
\delta_j ~W_{ij}(r)~ \; ,
\label{xi1est}
\ee where \be N_r \; = \; \sum_{i,j} W_{ij}(r) \ee is the number of
pairs of pixels at separation $r$ in the sampled region, and the
window function $W_{ij}(r)=1$ if pixels $i$ and $j$ are separated by
$|{\vec r}_i - {\vec r}_j| = r \pm \Delta r$, and 0 otherwise. For the
pairwise velocity we define as $\hat v_i$ the average velocity in the
$i$th pixel at position $\vec{r_i}$.  We can then use an equivalent
expression for the first estimator of $\vs$: 
\be {\hat v}^{(1)}(r) \;
= \; \sum_{i,j} (\vec{v_i}-\vec{v_j}) \cdot {\hat r}_{ij}~U_{ij}(r) \;
,
\label{vs1est}
\ee
where  
\be
{\hat r}_{ij} \; \equiv \; 
{{\vec{r_i}-\vec{r_j}} \over {|\vec{r_i}-\vec{r_j}|}} 
\ee
is a unit vector separating pixels $i$ and $j$ and the
sum is over all pairs of pixels, while
\be
U_{ij} \; \equiv \; {{(1+\delta_i)~(1+\delta_j)~W_{ij}(r)}
\over {N_r[1+\xi(r)]}} \; \; .
\ee
The second estimator uses all of the particle pairs rather than the
counts in pixel pairs:
\be
{\hat \xi^{(2)}}(r) \; =  \; {1\over {\langle n\rangle V_r}}
\sum_{i>j} ~W_{ij}'(r)~ -1 ,
\label{xi2est}
\ee where now the window function $W_{ij}'(r)=1$ if particles $i$ and
$j$ are separated by $|{\vec r}_i - {\vec r}_j| = r \pm \Delta r$, $\,
\langle n\rangle$ is the mean particle density and $V_r$ is the volume
of the spherical shell of radius $r$ and thickness $\Delta
r$. Typically, $\,\langle n\rangle\, V_r$ is estimated from a random
catalogue of particles, drawn from the same sample, 
because the effective
volume, containing the pairs separated by a distance $r \pm \Delta r$
might depend on the geometry.  However, our simulations use a large
and periodic box, with no boundaries and high densities (not affected
by shot-noise at the scales of interest). So the above expression gives a good and
quick estimator. The corresponding estimator for the pairwise velocity
is 
\be {\hat v}^{(2)}(r) \; = \; {\sum_{i>j} ~ (\vec{v_i}-\vec{v_j})
\cdot {\hat r}_{ij} ~W_{ij}'(r) \over {\sum_{i>j} ~W_{ij}'(r)}}
\label{vs2est}
\ee
where $v_i$ and $v_j$ are now individual particle velocities.
Note that this estimator does not depend on the effective
volume of the shell, $V_r$.

Both estimators agree reasonably well in our simulations. The first
set is more useful (faster to run) for large separations, as we can
reduce the resolution of the lattice and have a relatively small
number of pixels.  The second set is more adequate (faster to run) for
the small separations.

\subsection{The correlation function}

\begin{figure}
\centering
\centerline{\epsfxsize=9truecm 
\epsfbox{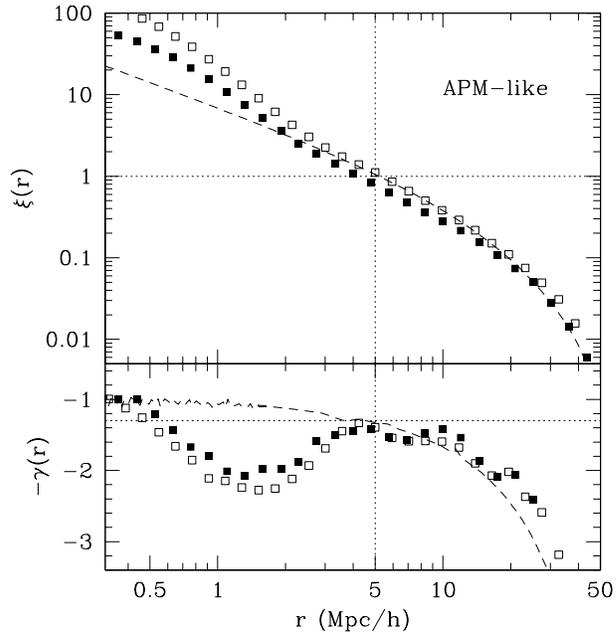}}
\caption[junk]
{The top panel shows the linear correlation function
(dashed line) and 
the measured non-linear $\xi(r)$, obtained from the APM-like 
simulations  with 
density parameters $\Omega_m=0.3$ (open squares) and $\Omega_m=1.0$ (full
squares). The bottom panel shows the corresponding logarithmic slope,
$\, -\gamma(r) = d \ln \xi /d\ln r$ for each of the three
curves from the top panel.
The vertical dotted line shows the separation
$r_o$, defined by the condition $\xi (r_o) = 1$ (top) and 
the separation $r_*$, at which
the non-linear $\gamma(r)$ curve crosses the linear one (bottom).}   
\label{x2sltest}
\end{figure}

The evolved, nonlinear correlation functions, measured
from simulations are shown in Figure \ref{x2sltest} (top panel).
The full squares correspond to the $\om = 1$ model, while the
open squares represent $\om = 0.3$. For comparison, we show the
linear correlation function (dashed line), 
scaled from some  
``initial'' redshift $z_o$ to
the present $(z = 0)$, following the linear theory expression
for the Einstein-de Sitter model, 
$\xi(r,0) = \xi(r,z_o)(1+z_o)^2$.
Nonlinear effects are more pronounced in the low
density model. This happens because of the well known
suppression of linear growth, which occurs at late times
$(\, z < 1/\om)$ in low density models and leads to the enhanced 
clustering on small scales relative to large scales.

Note however, that although the correlation functions differ
significantly in amplitude at separations
$r < 2 h^{-1}$Mpc, their slopes $\gamma(r)$ are
almost indistinguishable (Figure \ref{x2sltest}, bottom panel).

The particle resolution (the Nyquist wavelength $\propto N^{-1/3}$) of
the simulations used here is significantly lower than the resolution
of Virgo simulations, and  the noise in the measured $\xi(r)$ is further
amplified by differentiating over $r$. As a result, determining the
position of the inflection point $r_*$ directly from the $\gamma(r)$
curve alone is difficult. To overcome this problem, we identify $r_*$
by comparing the linear and nonlinear $\gamma(r)$ curves.  Taking
$r_*$ to be the separation at which the nonlinear slope drops below
the linear slope in Figure \ref{x2slapm}, as described earlier, we get
\be r_* \; \simeq \; 5 \,\Mlu \; \approx r_o \; ,
\label{simulated_r*}
\ee 
in excellent agreement with equation (\ref{precise_test}). 
Hence, the equality between $r_*$ and $r_o$ 
can probably be considered as a generic
outcome of gravitational dynamical evolution in a model where
galaxies trace the mass and the initial slope, $\, d\ln\xi/d\ln r$,
is a smooth decreasing function of the separation $r$. Such a
picture is also known as hierarchical clustering; an obvious
additional condition to make sure that small scale clumps collapse
before the large scale ones, is $\gamma > 0$, see e.g. LSS.

\subsection{The relative velocity}

\begin{figure}
\centering
\centerline{\epsfxsize=8.truecm 
\epsfbox{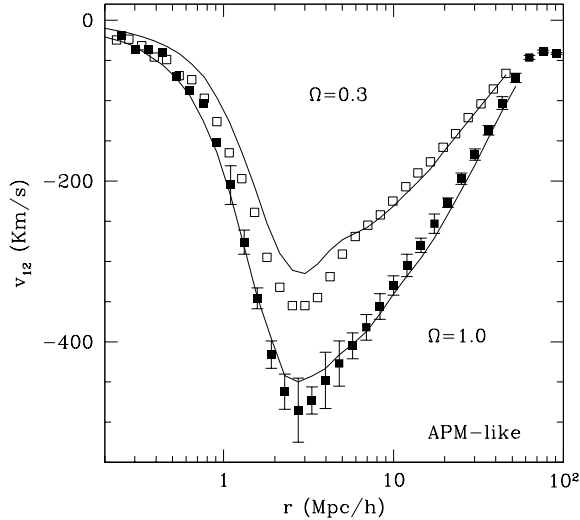}}
\caption[junk]
{The mean pairwise velocity $\vs(r)$,
measured from two sets of {\sc APM}-like 
simulations  with $\om=0.3$ 
(open squares) and $\om=1.0$ (full
squares), are compared with 
with an approximate 
analytical solution of the pair conservation equation
(eq.~[\ref{2nd-order}]; continuous lines).}
\label{v12c600}
\end{figure}

\begin{figure}
\centering
\centerline{\epsfxsize=9.truecm 
\epsfbox{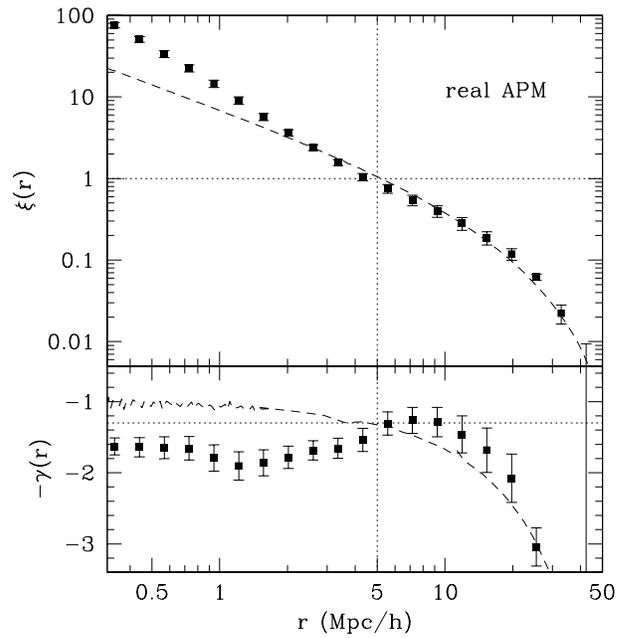}}
\caption[junk]{The spatial correlation function
of {\sc APM} galaxies (top panel, symbols
with errorbars), compared to the 
linear  theory {\sc APM-}like model,
described earlier (top panel, dashed line).
The bottom panel
shows the corresponding logarithmic 
slope, $\gamma(r)$.
The intersection of the two perpendicular
dotted lines marks the points where $\xi \simeq 1$ (top) and where
the non-linear slope crosses the linear one (bottom).}   
\label{x2slapm}
\end{figure}

In this section we describe N-body tests of the {\sc JSD} model for
the relative motions in pairs of galaxies.  We consider two models
with {\sc APM-}like initial spectra: an Einstein-de Sitter model and
an open model with $\om = 0.3$. For both models, the theoretical
predictions for the mean pairwise velocity, based on equation
(\ref{2nd-order}), are plotted in Figure \ref{v12c600} as continuous
lines. These predictions can be compared with N-body measurements,
shown as full squares for the $\om = 1$ model and as open squares for
$\om = 0.3$.  The agreement between the theoretical and experimental
$\vs(r)$ curves shows that our ansatz provides a good approximation of
the N-body results in the entire dynamical range probed for both
models. The mean and errors in the $\om = 1$ simulations come from the
mean and dispersion, obtained from five independent realizations 
of the APM-like model. For the open model ($\om
= 0.3$, open squares) we used only one realization, but the expected
sampling variance is expected to be the same. Indeed, the initial
$P(k)$ is identical in both cases.  Moreover, the long-wave tails of
the {\it final} power spectra (which determine the size of the
sampling error bars) are also identical because they are not affected
by the nonlinear evolution.

\section{Comparison with observations}

\subsection{The correlation function}

The measurements of $\xi_g(r)$, and $\, \gamma(r) \equiv - d\ln
\xi_g/d\ln r$, obtained from the angular correlations of galaxy pairs
in the {\sc APM} catalogue (Baugh 1996), are plotted in Figure
\ref{x2slapm}.  The top panel shows the two-point function (points
with error bars), and the linear theory curve, described in \S 3.3
(dashed line). The intersection of the two perpendicular dotted
lines marks the point $(\xi_g, r) = (1, r_o)$.  The
bottom panel of Figure \ref{x2slapm} shows the {\sc APM} $\gamma(r)$
as a function of the pair separation $r$. Note the remarkable
similarity between the empirical data and the characteristic peak in
the $\gamma(r)$ found in the simulations (Figure \ref{x2sltest}).  The
intersection of the two mutually perpendicular, dotted lines in the bottom
panel of Figure \ref{x2slapm} marks the crossing between the linear
model for $\gamma(r)$ (dashed line) and the nonlinear $\gamma(r)$
curve, determined from the {\sc APM} catalogue. The crossing occurs at
the separation $r \simeq 5\Mlu$, and to first approximation this scale
could be identified with $r_*$.  However, a closer inspection of our
Figure \ref{x2slapm} suggests that, given the error bars, the actual
position the peak could be shifted to the right, to a somewhat larger
separation.  Taking into account the error bars in Figure
\ref{x2slapm} as well as the uncertainties in the assumed linear
theory slope, we obtain \be r_* \; \simeq \; (6 \pm 1)~\Mlu \; , \ee
and \be r_o \; \simeq \; (5 \pm 1)~\Mlu \;.  \ee The slope at $r =
r_*$ is $\gamma_* \simeq -1.4$.  If we assume the linear bias model,
the relation $\, r_* \; \approx \; r_o \, b^{-2/\gamma_*}$ gives \be b
\; = \; 1.11 \pm 0.22 \;
\label{b-constraint}
\ee
at one-sigma statistical significance level.

\subsection{The relative velocity}

We will now apply the second of the two proposed tests of biasing:
the relative velocity test. We will compare the mean pairwise velocity,
predicted by assuming that the {\sc APM} galaxies trace the mass with
the pairwise velocity, measured directly from a peculiar velocity --
distance survey. 

\begin{figure}
\centering
\centerline{\epsfxsize=9.truecm 
\epsfbox{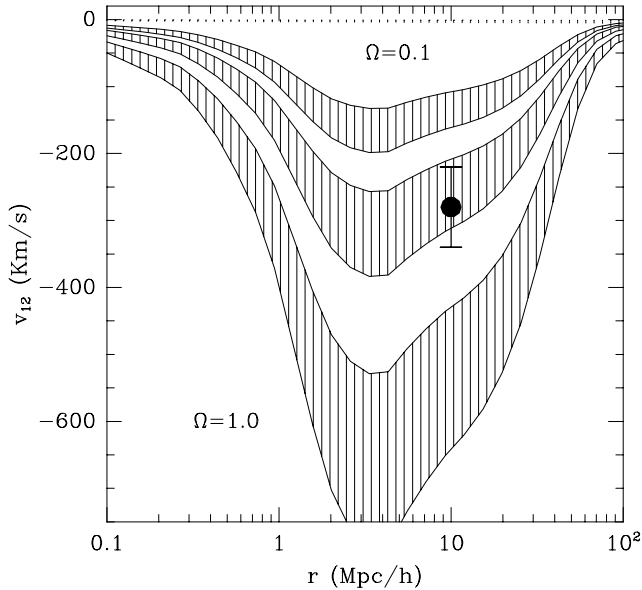}}
\caption[junk]{Predictions for
the mean pairwise velocity $v_{12}$, based on the assumption
that the {\sc APM} galaxies trace the mass. The shaded regions
correspond to $20\%$ uncertainties in the strength of clustering.   
We consider three values of $\Omega_m=1.0,0.3,0.1$, top to
bottom (as labeled). The point with error bars corresponds to
a direct measurement from the Mark III survey 
\cite{rj00}.}
\label{v12apm}
\end{figure}

Figure \ref{v12apm} shows $\vs(r)$ curves, predicted by
equation (\ref{2nd-order}) (continuous
lines) for three different values of $\Omega_m$, from bottom to top
$\Omega_m=1.0,0.3,0.1$, respectively. To calculate
$\vs(r)$, we have used $\xi(r)$, estimated from the
{\sc APM} survey under the assumptions $\xi_g = \xi$ and $R = 1$. 

Before making the comparison, we must overcome the following problem.
The survey  has a significant
depth, with the mean redshift of $z \simeq 0.15$ while
the measured $\vs(r)$ corresponds to the present time $(z = 0)$.
To evolve $\xi(r,z)$ from $z \simeq 0.15$ to $z=0$, we need to make some
additional model assumptions. 
Gazta\~naga (1995) has shown that for this redshift range,
the uncertainties in the details of dynamical evolution of $\xi$
are small. In particular, choosing an incorrect value
for $\om$ can affect $\xi$ at most at the several
per cent level (for $\om$ ranging from 1 to 0). Adding this
to other possible sources of errors, such as uncertainties regarding
the redshift evolution of the galaxy number density and sampling
and selection fluctuations, Gazta{\~n}aga (1995) estimates
that the resulting relative uncertainty in the amplitude
of $\xi_g$ is $ \lsim 20\%$. According to his analysis,
the present $(z = 0)$ amplitude
of the rms fluctuation of the {\sc APM}
galaxy counts, measured in spheres of radius
of $8~h^{-1}$Mpc, is
$\, 1.1 \lsim \se^{APM} \lsim 0.9$. 

To be conservative, for each value of $\om$, we plot the predicted
$\vs(r)$ curves for two values of $\se$, differing by $20\%$.  The
resulting prediction for each value of $\om$ is therefore an
area rather than a single $\vs(r)$ curve (see Figure 4).  The lower
boundary of each shaded area assumes $\se = 1.1$ while the upper
boundary was calculated by assuming $\se = 0.9$.  A direct measurement
at $r = 10~h^{-1}$ from the Mark III galaxy peculiar velocity survey
(Willick et al. 1997) gives \cite{rj00} 
\be \vs \; = \;
-280 \pm 60 \; {\rm km/s} \; .
\label{direct}
\ee
It is reassuring that this value, plotted in Figure 4,
overlaps with the shaded area, corresponding to $\om = 0.3$,
because it agrees with ranges for $\om$ and $\se$ obtained
by Juszkiewicz et al. from the analysis of the Mark III
survey alone. Their one-sigma constraints are $\, \om =
0.35^{+0.35}_{-0.25}$ and $\se \ge 0.7$, and the analysis 
assumes that the correlation function for the mass is
well approximated by a pure power law with $\gamma = 1.75$.
 
From the agreement
between the predicted and observed
value of the mean pairwise 
velocity, we conclude that the Mark III and {\sc APM}
data, taken together, are consistent with the hypothesis
that the {\sc APM} galaxies trace the mass, $b \approx R \approx 1$,
while the density parameter is low, $\om \approx 0.3$. 

\subsection{How can biasing affect $\vs(r)$ ?}

Apart from leading to predictions which are confirmed observationally, 
the ``what you get is what you see'' hypothesis has another
important advantage: simplicity. Once $\xi(r)$ is estimated
from observations, a family of $\vs(r,\om)$ curves
can be calculated directly from
equation (\ref{2nd-order}) for any given range of values
of the density parameter. This simplicity will immediately
go away if we allow scale-dependent, stochastic and nonlinear
biasing. A frank answer to the question posed in the heading
of this subsection would have to be ``Only God (of biasing) knows''. 
Predicting $\vs(r)$ without resorting
to massive numerical simulations would be simply impossible.
We can get the idea of what is in store by considering only
the leading order term in the perturbative expansion for
$\vs(r)$ at large separations \cite{moriond},
\be
\vs(r) \; = \; -{\textstyle {2\over 3}}\,Hrf(\om)\,\xb_{g\rho}(r) \; ,
\label{biased_v12}
\ee
where 
$\,\xb_{g\rho}(r) = (3/r^3)\,\int_0^r\,\xi_{g\rho}(x)x^2dx\,$ 
is the galaxy-mass cross-correlation function,
\be
\xi_{g\rho}(r) \; = \; \langle\delta(0)\delta_g(\vec{r})\rangle \; ,
\ee 
averaged over a sphere of radius $r$. The function
$\xi_{g\rho}$ describes the cross-correlations between the mass density 
and the density of galaxies in the velocity field survey, which
in the case considered here would be the Mark III catalogue.

To make progress in our analysis, we will now generalize the
definition of the stochasticity parameter introduced as
a normalized cross-correlation of two random fields, $\delta$
and $\delta_g$, measured at the same position in space.
Instead, we will consider a cross-correlation of the
same two fields measured at two different positions
in space, separated by distance $r$. Our old equation 
(\ref{bullshit_coefficient}) is replaced by
\be
R(r) \;\; = \;\; {\xi_{g\rho}(r)\over\sqrt{\xi(r)\xi_g(r)}} \;\; .
\ee 
Let us make another simplifying assumption, that 
$\, b^2(r) \equiv \xi_g(r)/\xi(r)$,
as well as $R$, are separation-independent. 
Equation (\ref{biased_v12}) becomes
\be
\vs(r)\; = \; - {\textstyle{2\over3}}\,
f(\om)\,Hr \, R\,b\,\xb(r) \;\;.
\label{baroque}
\ee
The expected relative pairwise velocity can now be related to
the {\sc APM} data by substituting
\be
\xb(r) \; = \; {3\over \bapm^2 \, r^3}\, 
\int_0^{r}\, \xi_g(x)x^2dx \; ,
\label{b_APM}
\ee
where $\,\bapm^2 = \xi_g(r)/\xi(r)$ and $\xi_g$ is the
{\sc APM} galaxy correlation function. 
In case of trouble in predicting the correct $\vs(r)$ curve,
we now have three essentially free parameters which can be readjusted.
This is only the tip of the iceberg, as we have ignored
nonlinear dynamics as well as the scale-dependence of $b$ and $R$.

For the linear bias model, the predictions are in clear
conflict with the data unless $b$ is close to unity.
After setting $R = 1$, we get
$\vs(r) \propto b$. Then, if the biasing factor for
spiral galaxies is, as usually assumed $b \approx 1$,
our predictions for $\vs(r)$ in the linear regime ($r \gsim 10 \Mlu$)
should be similar to the unbiased predictions already
plotted in Figure (\ref{v12apm}). If the biasing factor
for the ellipticals is significantly different, say, $b \approx 2$,
the elliptical subsample of the Mark III
data should give estimates of $\vs$ which differ from
the estimates from the spiral sample by the same factor of two.
Meanwhile the estimates from the appropriate subsamples 
in the real data are indistinguishable \cite{rj00}.
Hence, just as in case of the shape of the {\sc APM} correlation
function, considered above, the deterministic linear biasing
model is inconsistent with observations.

We can summarize the last two subsections as follows. The prediction for
$\vs(r)$, based on the assumption that the {\sc APM} galaxies
trace the mass passes our test as it agrees with the velocity,
estimated from the Mark III data. The simplest prescription
of biasing fails the test. More complicated prescriptions
can probably be made to pass, which is not surprising, given the
number of free parameters.

\section{Summary and discussion}

A quarter of a century ago, Gott and Rees predicted that gravity
should leave its mark on the shape of the galaxy autocorrelation
function: a ``shoulder'', or steepening of the slope of the
correlation function should appear near the separation where $\xi$
passes through unity. At the time, biasing was unheard of, and Gott
and Rees (1975) assumed $\xi = \xi_g$. Recently, in another context,
JSD have studied the the $\xi = 1$ boundary in the evolution of the
mass correlation function, using results from Virgo simulations. They
found that the ``shoulder'' is actually an inflection point, occurring
at a well defined separation $r_*$.  In all four {\sc CDM}-like models
they studied, the nonlinear transition looked strikingly similar: the
inflection occurred at almost the same separation as that of the
nonlinear transition: $r_* \approx r_o$, where $r_o$ corresponds to
$\xi = 1$.  Here we have tested the degree of universality of their
results by widening the range of models considered. Our additional
objective was to study the range of validity of an approximate
solution of the pair conservation equation, proposed by JSD to study
the nonlinear evolution of the relative velocity of particle pairs at
a fixed separation, $\, \vs(r)$.  We used N-body simulations, with
{\sc APM}-like initial conditions, with two different values of the
density parameter: $\om = 1$ and 0.3. The {\sc APM}-like initial power
spectra differ significantly from all of the {\sc CDM}-like spectra,
considered earlier by JSD. Moreover the spectra of the latter kind
appear as more realistic to us because they can reproduce observations
without resorting to scale-dependent biasing.  Our {\sc APM}-like
simulations are in excellent agreement with earlier results,
confirming the validity of the JSD ansatz for $\vs(r)$ and the
conjecture that the appearance of the shoulder in the correlation
function near the $\xi = 1$ transition is a feature of gravitational
dynamics rather than a peculiarity of a particular set of initial
conditions.

Using these results, we proposed two tests of the hypothesis that
galaxies trace mass. The first of the tests is based on an obvious
idea, that if $\xi(r) = \xi_g(r)$, the galaxy correlation function
near $\xi_g =1$ should exhibit properties similar to those of the
matter correlation function. We examined the behavior of the
correlation function, derived from the {\sc APM} catalogue and found
exactly the same features we knew earlier from N-body simulations, in
particular the agreement between the two characteristic scales, $r_*
\approx r_o$. It is difficult to imagine how such an agreement could
happen by a mere coincidence, which would have to be the case if
$\xi_g$ is unrelated to $\xi$. The agreement between the two
characteristic scales can be used to constrain the linear biasing
factor for the {\sc APM} catalogue to be within $20\%$ of unity. This
constraint agrees with an earlier limit, obtained from measurements of
the three-point correlation function from the {\sc APM} survey
(Gazta{\~n}aga 1994, Frieman \& Gazta{\~n}aga 1999).

The second test confronts the $\vs$, predicted by assuming
that the {\sc APM} galaxies are unbiased tracers of mass with
direct measurements of $\vs$. The results are again consistent
with $b \approx 1$ and a low density parameter, $\om \approx 0.3$,
in agreement with the limits obtained from the velocity data alone
\cite{rj00}.

We are impressed how well the observations are
reproduced by the simple calculations
based on the assumption that galaxies follow
the mass distribution, at least on large (weakly non-linear)
scales. We are unable to constrain biasing
models with a large number of free parameters, but 
their predictive power is questionable and one may ask:
are such models falsifiable and therefore worth constraining?   

Our results are by no means final, they are also less rigorous than
one could wish because we are limited by the accuracy of the
present observational data. New generation of catalogues promise
a dramatic improvement on this front in the near future
(for an excellent collection of reports on the state of the
art in this field, see Colombi et al. 1998).

\section{Acknowledgements}

One of the co-authors (RJ) who was more lucky than the other
and who attended this excellent Meeting would like to thank Jim Fry
for making it happen. We also thank Carlton Baugh for providing us with his
{\sc APM}-like simulations as well as his estimate of $\xi_g(r)$,
based on the {\sc APM} survey.
RJ thanks
Ruth Durrer for important discussions regarding the inflection point
in $\xi(r)$ and for her hospitality at the University of Geneva.  This
work was supported by a collaborative grant between the Polish Academy
of Science and the Spanish Consejo Superior de Investigaciones
Cientificas.  We also acknowledge support by grants from the Polish
Government (KBN grant No. 2.P03D.01719), the Swiss
Tomalla Foundation, and from IEEC/CSIC and DGES(MEC) (Spain), project
PB96-0925.


\def\refe {\par \hangindent=.7cm \hangafter=1 \noindent}
\def\apj {ApJ}
\def\na {Nature,}
\def\aap {A\&A}
\def\apjs{ApJS}
\def\mn {MNRAS}

\end{document}